\documentclass[aps,prl,twocolumn,superscriptaddress]{revtex4-1}

\usepackage{verbatim}
\usepackage{amsmath}
\usepackage[utf8]{inputenc}
\usepackage[pdftex]{graphicx}
\usepackage[thinspace]{SIunits}
\usepackage[colorlinks=true,urlcolor=blue,linkcolor=blue,citecolor=blue]{hyperref}
\usepackage{xcolor}
\usepackage[T1]{fontenc}
\usepackage{placeins}  
\usepackage{nicefrac} 
\usepackage{braket}

\begin{document}
\title{Monolithically integrated single quantum dots coupled to bowtie nanoantennas}
%
%
%
%
\author{A.~A.~Lyamkina}\thanks{These two authors contributed equally.}
\affiliation{A. V. Rzhanov Institute of Semiconductor Physics SB RAS, Pr. Lavrentieva 13, 630090 Novosibirsk, Russia}
\affiliation{Novosibirsk State University, Pirogova 2, 630090 Novosibirsk, Russia}
\author{K.~Schraml}\thanks{These two authors contributed equally.}
\affiliation{Walter Schottky Institut and Physik Department, Technische Universit\"at M\"unchen, Am Coulombwall 4, 85748 Garching, Germany}
\author{A.~Regler}
\affiliation{Walter Schottky Institut and Physik Department, Technische Universit\"at M\"unchen, Am Coulombwall 4, 85748 Garching, Germany}
\affiliation{Technische Universit\"at M\"unchen, Institute for Advanced Study, Lichtenbergstrasse 2a, 85748 Garching, Germany}
\author{M.~Schalk}
\affiliation{Walter Schottky Institut and Physik Department, Technische Universit\"at M\"unchen, Am Coulombwall 4, 85748 Garching, Germany}
\author{A.~K.~Bakarov}
\affiliation{A. V. Rzhanov Institute of Semiconductor Physics SB RAS, Pr. Lavrentieva 13, 630090 Novosibirsk, Russia}
\affiliation{Novosibirsk State University, Pirogova 2, 630090 Novosibirsk, Russia}
\author{A.~I.~Toropov}
\affiliation{A. V. Rzhanov Institute of Semiconductor Physics SB RAS, Pr. Lavrentieva 13, 630090 Novosibirsk, Russia}
\author{S.~P.~Moshchenko}
\affiliation{A. V. Rzhanov Institute of Semiconductor Physics SB RAS, Pr. Lavrentieva 13, 630090 Novosibirsk, Russia}
%
%
\author{M.~Kaniber}\email{michael.kaniber@wsi.tum.de}
\affiliation{Walter Schottky Institut and Physik Department, Technische Universit\"at M\"unchen, Am Coulombwall 4, 85748 Garching, Germany}
%
%
%
%
\begin{abstract}
Deterministically integrating semiconductor quantum emitters \cite{bimberg1999quantum} with plasmonic nano-devices \cite{Lal2007, Schuller2010} paves the way towards chip-scale integrable \cite{Zia2006}, true nanoscale quantum photonics technologies \cite{OBrien2010}. For this purpose, stable and bright semiconductor emitters \cite{Hoang2015} are needed, which moreover allow for CMOS-compatibility \cite{Ozbay2006} and optical activity in the telecommunication band \cite{Goldmann2014}. Here, we demonstrate strongly enhanced light-matter coupling of single near-surface ($<10\,nm$) InAs quantum dots monolithically integrated into electromagnetic hot-spots of sub-wavelength sized metal nanoantennas. The antenna strongly enhances the emission intensity of single quantum dots by up to $\sim16\times$, an effect accompanied by an up to $3.4\times$ Purcell-enhanced spontaneous emission rate. Moreover, the emission is strongly polarised along the  antenna axis with degrees of linear polarisation up to $\sim85\,\%$. The results unambiguously demonstrate the efficient coupling of individual quantum dots to state-of-the-art nanoantennas. Our work provides new perspectives for the realisation of quantum plasmonic sensors \cite{Lee2016}, step-changing photovoltaic devices \cite{Atwater2010}, bright and ultrafast quantum light sources \cite{Hoang2016} and efficent nano-lasers \cite{Bergman2003}.
\end{abstract} 
%
%
\maketitle
%
The field of nanoplasmonics \cite{maier2007plasmonics} has already demonstrated outstanding potential to tailor and enhance electromagnetic fields on sub-wavelength lengthscales \cite{Schuller2010}. As such it represents the most promising route to interface state-of-the-art electronics with true nano-photonic devices on the same chip \cite{Ozbay2006}. To this end, the study, optimisation and integration of nano-scale plasmonic components, such as antennas \cite{Fromm2004} and waveguides \cite{Zia2006}, on high-quality semiconductor substrates \cite{Schraml2014} is essential in order to prove their applicability in real-world applications. Monolithically integrated, self-assembled quantum dots \cite{bimberg1999quantum} exhibit outstanding electrical and optical properties, they do not suffer from bleaching or blinking and have near-unity internal quantum efficiencies. These properties stem from the efficient decoupling from environmental perturbations in the solid-state matrix material and distinguish self-assembled quantum dots from alternative quantum emitters, such as nitrogen vacancy centres \cite{Schell2011}, colloidal nano-crystals \cite{Hoang2015}, single molecules \cite{Kinkhabwala2009} or fluorescent dyes \cite{Akselrod2014}. Lithographically defined plasmonic dimer antennas, such as bowties \cite{Fromm2004}, are most prominent amongst the zoo of metallic nanoparticles since they simultaneously provide strong light confinement in sub-wavelength sized hot-spots, large-range spectral tunability and facilitate electrical access \cite{Prangsma2012} and full control of the emission polarisation \cite{Biagioni2009}. As a result, the quantum dot coupled nanoantennas offer new perspectives to probe light-matter-couplings and cavity quantum electrodynamics (cQED) effects  beyond the point-dipole approximation \cite{Andersen2011}.

%
\begin{figure*}
\includegraphics{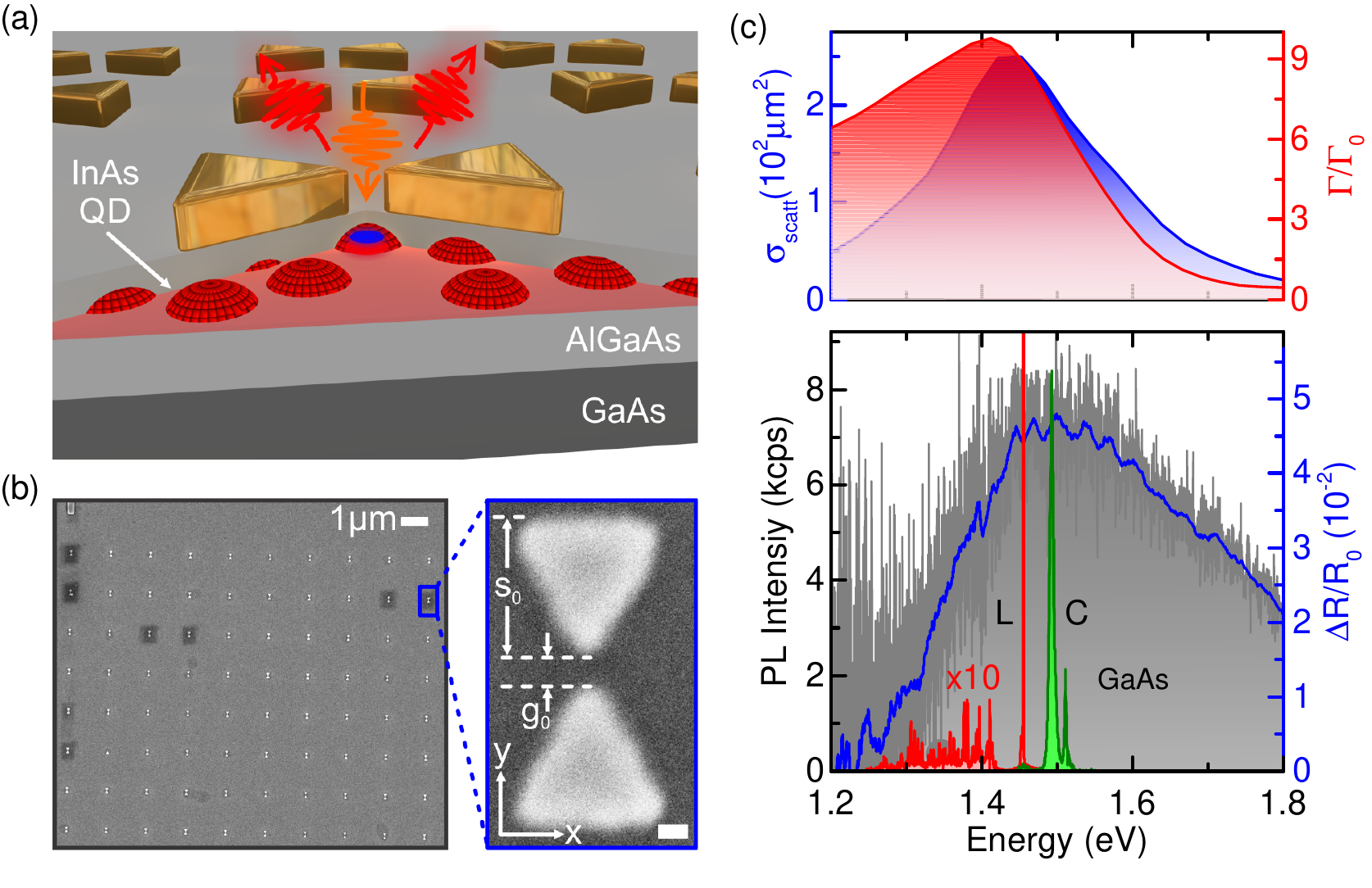}
\renewcommand{\figurename}{Figure}
\caption{\label{fig1}
\textbf{Sample layout, structural and optical characterisation:} (a) Schematic illustration of the monolithically integrated quantum dot-nanoantenna system (b) Scanning electron microscopy image of the bowtie nanoantenna array (left) and an individual bowtie nanoantenna (right) with $s=87\,nm$ and $g=26\,nm$. Scale bar, $20\,nm$. (c) Top: Simulated scattering cross-section $\sigma_{scatt}$ and normalised decay rate $\Gamma/\Gamma_0$ of a bowtie nanoantenna with nominal $s_0=90\,nm$ and $g_0=15\,nm$ in blue and red, respectively. Bottom: Measured quantum dot spectrum (L: laser, C: carbon impurities) and differential reflectivity of a bowtie nanoantenna with $s=87\,nm$ and $g=26\,nm$ in red/green and grey, respectively. Blue curve shows a 50-point-smoothed average.
}
\end{figure*}
%

In this Letter, we coupled individual quantum dots to plasmonic nanoantennas to form a novel cQED-system schematically illustrated in figure~\ref{fig1} (a), which consists of near-surface ($d\sim10\,nm$) InAs/AlGaAs quantum dots \cite{Adlkofer2002}  and lithographically defined triangular bowtie nanoantennas \cite{Fromm2004} arranged in a square array (left scanning electron microscopy image in figure~\ref{fig1} (b)) with a lattice constant $a=1.5\,\mu m$. A typical nanoantenna with triangle size $s=87\pm 3\,nm$ and feed-gap size $g=26\pm 3\,nm$ as shown in the right scanning electron microscopy image in figure~\ref{fig1} (b). The self-assembled InAs quantum dots are grown by solid-source molecular beam epitaxy and embedded into high-band gap AlGaAs (see Methods) in order to suppress carrier tunnelling from the quantum dots to trap states at the nearby sample surface. In the top panel of figure~\ref{fig1} (c), we present simulations for a bowtie nanoantenna ($s_0=90\,nm$ and $g_0=15\,nm$) of the scattering cross-section $\sigma_{scatt}$ (blue curve) and the normalised decay rate $\Gamma/\Gamma_0$ (red curve), where $\Gamma$ and $\Gamma_0$ denote the decay rate of a point-dipole with and without the proximal bowtie nanoantenna, respectively. We observe a $\Delta\sim30\,meV$ red-shift of the maximum $\Gamma/\Gamma_0$ with respect to the maximum $\sigma_{scatt}$, which is attributed to the difference between the near- and the far-field coupling \cite{Zuloaga2011}. In the bottom panel of figure~\ref{fig1} (c), we show the experimentally determined differential reflectivity spectrum (raw data in grey, 50-point-smoothed average in blue) of the according nanoantenna that reveals the surface plasmon resonance and find a good agreement with the simulated $\sigma_{scatt}$ \cite{Kaniber2016}. Moreover, the red and green curves show typical photoluminescence spectra of the quantum dot ensemble and the GaAs substrate recorded spatially displaced from the nanoantennas. A good spectral overlap clearly exists between the quantum dot emission and the near-field plasmonic mode.

%
\begin{figure*}
\includegraphics{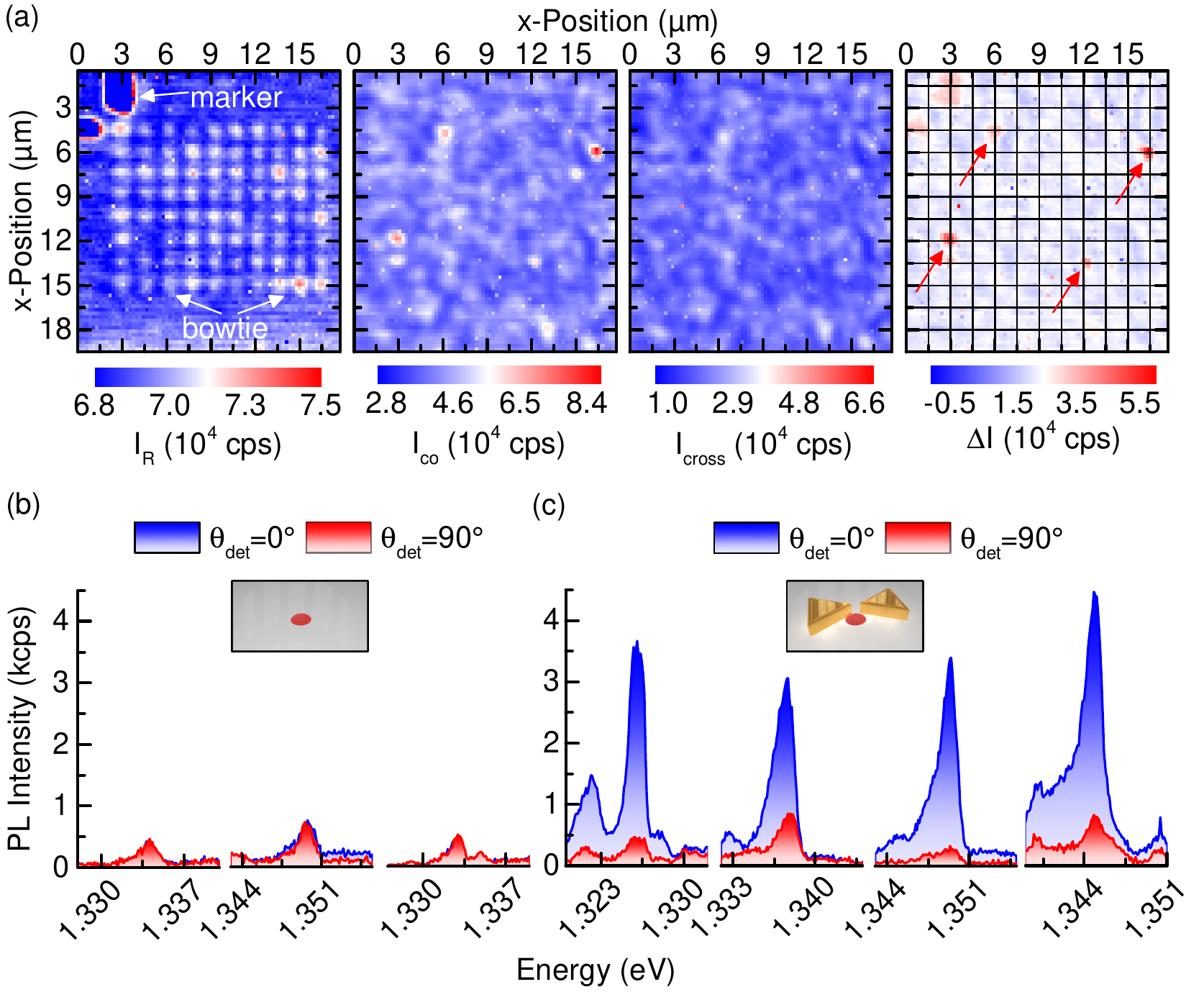}
\renewcommand{\figurename}{Figure}
\caption{\label{fig2}
\textbf{Enhanced luminescence from coupled quantum dot-nanoantenna systems:} (a) Spatially resolved laser reflectivity, co-, cross-polarised and differential quantum dot emission from left to right, respectively. (b), (c) Photoluminescence spectra of uncoupled and coupled quantum dots for co- and cross-polarised emission in blue and red, respectively.
}
\end{figure*}
We identify quantum dots that are spatially coupled to the bowtie nanoantenna using spatially resolved photoluminescence spectroscopy. Hereby, we simultaneously record the quantum dot emission and the partly reflected excitation laser that is co-polarised along the main antenna axis (i.e. y-axis in figure~\ref{fig1} (b)). The latter enables us to determine the spatial position of the bowtie nanoantennas with an accuracy of $250\,nm$ and, thus, to correlate it with the quantum dot emission. In figure~\ref{fig2} (a), we present experimental data of the laser reflectivity $I_R$, the quantum dot emission in co-polarised $I_{co}$ and cross-polarised $I_{cross}$ detection geometry and the calculated differential signal $\Delta I=I_{co}-I_{cross}$ from left to right, respectively. The $\Delta I$-map exhibits four spatially distinct positions (highlighted by arrows) with clearly enhanced responses which are spatially well correlated to the bowtie nanoantenna positions extracted via the $I_R$-map (highlighted by the grid). Spectrally resolved photoluminescence recorded from three uncoupled (i.e. reference) quantum dots located outside the bowtie nanoantenna array and of those four coupled quantum dot-nanoantenna systems are shown for co- (red) and cross- (blue) polarised detection in figure~\ref{fig2} (b) and (c), respectively. For the antennas with quantum dots we observe an up to $\sim 16\times$ enhanced photoluminescence intensity when comparing coupled and uncoupled quantum dots in a co-polarised detection geometry. We attribute this enhancement to a pronounced coupling of the emitter to the plasmonic mode, resulting in a combination of excitation enhancement \cite{Pfeiffer2010}, Purcell-enhanced emission \cite{Akselrod2014} and spatial redistribution \cite{Curto2010}.

%
%
\begin{figure*}
\includegraphics{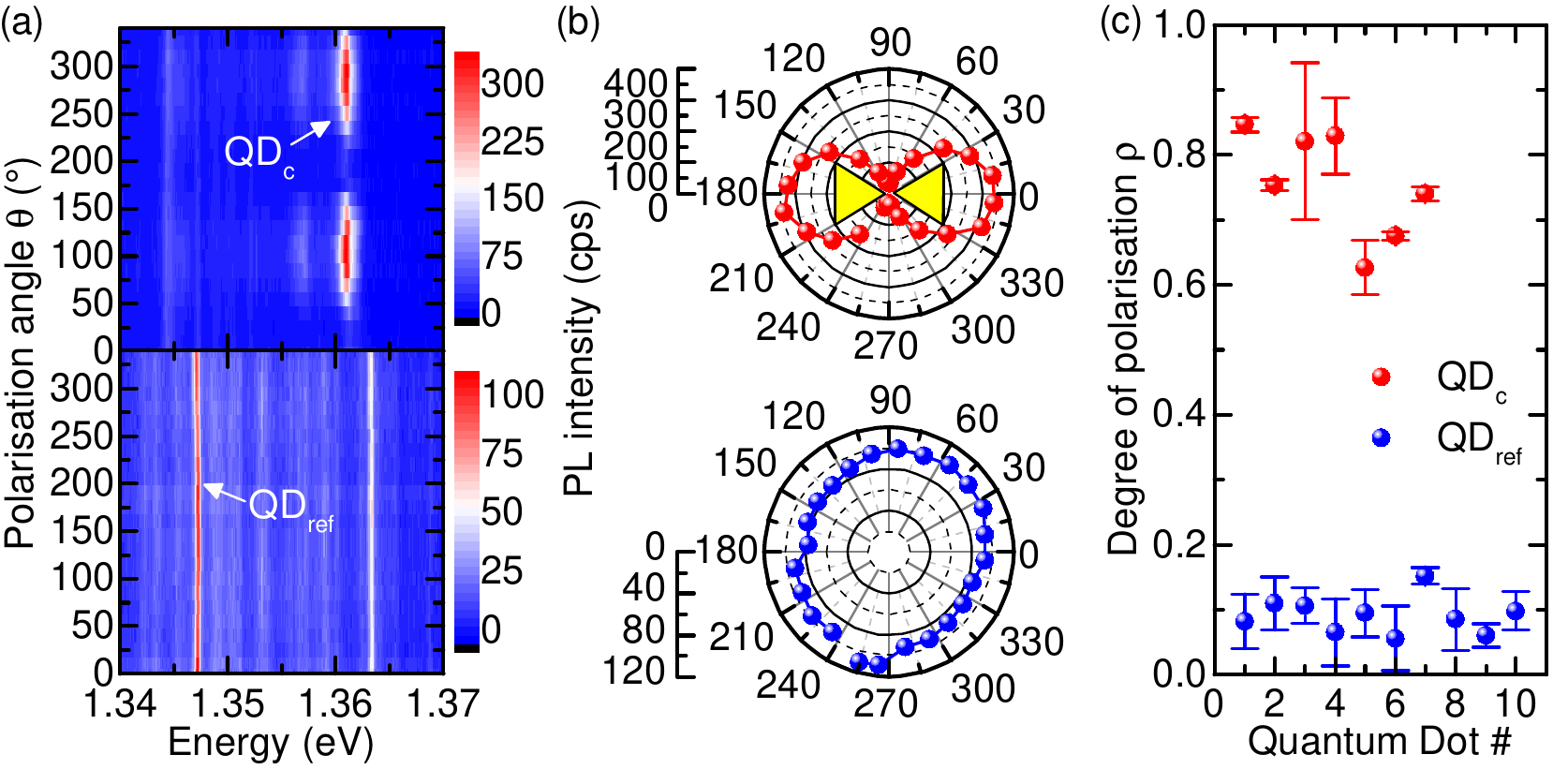}
\renewcommand{\figurename}{Figure}
\caption{\label{fig3}
\textbf{Strongly linear polarised emission from coupled quantum dot-nanoantenna systems:} (a) Polarisation-resolved photoluminescence for a coupled quantum dot-nanoantenna system (top) and an uncoupled quantum dot (bottom). (b) Corresponding peak intensities of a coupled quantum dot-nanoantenna system (top) and an uncouped quantum dot (bottom) in a polar-plot representation. (c) Degree of linear polarisation $\rho$ for coupled quantum dot-nanoantenna systems (red) and uncoupled quantum dots (blue).
}
\end{figure*}
In order to explore the coupling mechanisms between the quantum dots and the bowtie nanoantennas, we present in figure~\ref{fig3} (a) and (b) polarisation-resolved photoluminescence studies for a coupled quantum dot-nanoantenna system (top) and an uncoupled quantum dot (bottom) in a false-color and polar-plot representation, respectively. We observe strongly polarised emission from the coupled quantum dot-nanoantenna systems along the main antenna axis, yielding degrees of linear polarisation $\rho=(I_{max}-I_{min})/(I_{max}+I_{min})$ of up to $\rho_{coupled}=85$\%. In strong contrast, for the reference quantum dots we observe predominantly unpolarised emission with $\rho_{ref}=9\pm3$\% as shown in figure~\ref{fig3} (c). The results presented show that the degree of polarisation and the according polarisation angle of self-assembled InAs quantum dots are determined by the coupling to the highly polarised dipolar fields of the bowtie nanoantennas. Our results are found to be in excellent agreement with numerical simulations and, moreover, are in accord with studies performed on colloidal nanocrystals \cite{Curto2013}. 

%
%
\begin{figure*}
\includegraphics{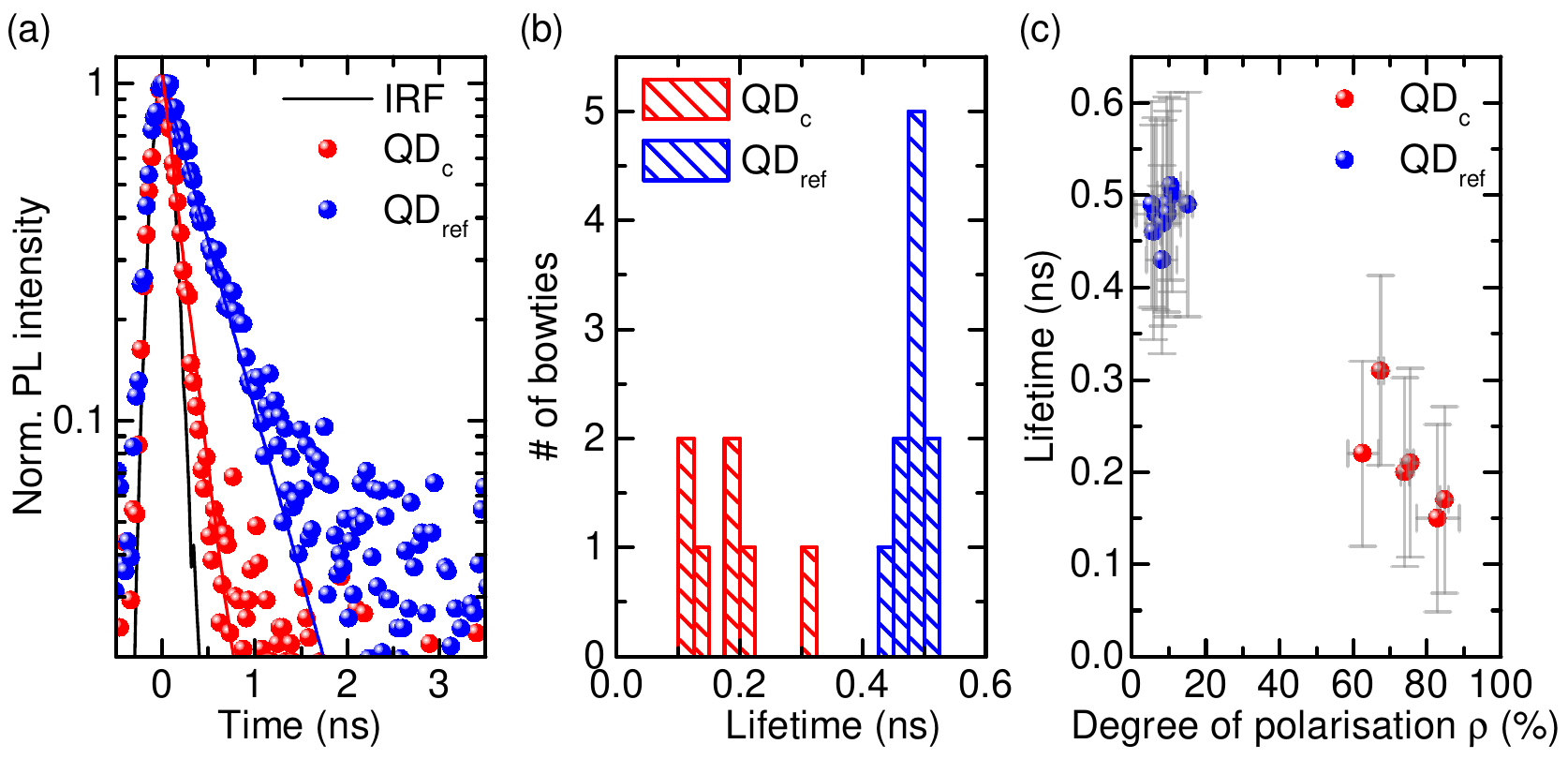}
\renewcommand{\figurename}{Figure}
\caption{\label{fig4}
\textbf{Purcell-enhanced emission from coupled quantum dot-nanoantenna systems:} (a) Time-resolved photoluminescence transients of a coupled quantum dot-nanoantenna system (red), an uncoupled quantum dot (blue) and the instrument response function (black). (b) Histogram of the spontaneous emission lifetimes for coupled quantum dot-nanoantenna systems (red) and uncoupled quantum dots (blue). (c) Correlation between the spontaneous emission lifetime and the degree of linear polarisation.
}
\end{figure*} 
We continue to study the contribution of the Purcell-enhancement to the overall intensity increase of coupled quantum dot-nanoantenna systems by applying time-resolved photoluminescence spectroscopy and compare our results to uncoupled quantum dots. In figure~\ref{fig4} (a), we present the time-resolved transients of a coupled quantum dot-nanoantenna system (red) and a reference quantum dot (blue) in a semi-logarithmic representation. We observe a significantly reduced spontaneous emission lifetime for the coupled quantum dot-nanoantenna system of $\tau_{coupled}=0.18\pm0.1\,ns$ (time-resolution limited), as compared to the uncoupled dot ($\tau_{uncoupled}=0.44\pm0.1\,ns$). In general, we find average lifetimes of coupled and uncoupled quantum dots of $\widetilde{\tau}_{coupled}=0.21\pm0.1\,ns$ and $\widetilde{\tau}_{uncoupled}=0.48\pm0.1\,ns$, respectively, as summarised in figure~\ref{fig4} (b), which we attribute to varying spatial displacements between the quantum dot positions and the antenna hot-spots \cite{Pfeiffer2014}. The clear decrease in spontaneous emission lifetime accompanied with the pronounced increase of emission intensity unambiguously demonstrates a modification of the radiative emission dynamics via the Purcell-effect \cite{Purcell1946}, giving rise to Purcell-factors $F_{P}>3.4$. Moreover, we observe a clear correlation between the measured spontaneous emission lifetimes and the corresponding linear degrees of polarisation as shown in figure~\ref{fig4} (c), indicating that pronounced modifications of the spontaneous emission lifetimes are accompanied by highly linear polarised emission. 

%
%
%
In summary, we demonstrated pronounced light-matter-coupling of bright ($\sim16\times$ enhanced) and fast ($<200\,ps$), monolithically integrated, semiconductor InAs quantum dots positioned $10\,nm$ below small feed-gap ($25\,nm$) lithographically defined Au bowtie nanoantennas, resulting in Purcell-factors $F_P>3.4$. The combination of such exquisite quantum emitters with lithographically engineered metallic nanoantennas \cite {Fromm2004, Pfeiffer2010} enables intrinsic electrical connectivity \cite{Prangsma2012} and, thus, expands the currently available portfolio of experimental techniques towards photocurrent spectroscopy and Stark tuning \cite{Fry2000}. Moreover, the studied hybrid system facilitates the extension to so-called cross-antennas \cite{Biagioni2009}, enabling full control of emission polarisation and, thus, represents a novel platform for advanced resonant fluorescence spectroscopy \cite{Flagg2009}. Further potential for optimising the light-matter-interaction is envisioned by deterministic quantum dot-antenna positioning \cite{Pfeiffer2014}, by engineering the capping of the semiconductor-heterostructure \cite{Zhang2015} and the in-situ epitaxial growth of monocrystalline metals \cite{Smith1996} that most likely also leads to a surface passivation of the sample. Moreover, the presented InAs quantum dot-nanoantenna system is a natural candidate to study in detail light-matter-interactions in mesoscopic quantum systems \cite{Andersen2011}, since the confined electromagnetic field is on the order or even smaller as compared to the spatial extensions of the electronic wavefunction. This would be expected to result in strongly modified optical selection rules and, thus, enables further enhancement of spontaneous emission, an essential requirement for applications in diverse scientific fields such as quantum information processing, optical quantum computing or light harvesting devices.

%
%
%
\hfill\break
\textbf{Methods}

\textbf{Sample fabrication.} The semiconductor heterostructure was grown by molecular beam epitaxy on a (001) GaAs substrate. After deposition of $500\,nm$ GaAs, we grew $50\,nm$ of $Al_{0.3}Ga_{0.7}As$. Self-assembled InAs quantum dots were formed at a temperature of $485\,^{\circ} C$ during a growth time of $35\,s$. After a growth interruption of $70\,s$ the quantum dots were capped by $10\,nm$ of $ Al_{0.3}Ga_{0.7}As$ to make them optically active.

The plasmonic antennas investigated were defined on semi-insulating (001) GaAs wafers. After cleavage, the samples were flushed with acetone and isopropanol (IPA). In order to get a better adhesion of the e-beam resist, the samples were put on a hot plate ($170\,^{\circ} C$) for $300\,s$. An e-beam resist (Polymethylmethacrylat 950 K, AR-P 679.02, ALLRESIST) was coated at $4000\,rpm$  for $40\,s$ at an acceleration of $2000\,rpm/s$ and baked out at $170\,^{\circ} C$ for $300\,s$, producing a resist thickness of $70\pm5\,nm$. The samples were illuminated in a Raith E-line system using an acceleration voltage of $30\,kV$ and an aperture of $10\,\mu m$. A dose test was performed for every fabrication run, as this crucial parameter depends on the varying e-beam current, with typical values ranging between $1200\,\mu C/cm^2$ and $1500\,\mu C/cm^2$. All samples were developed in Methylisobutylketon diluted with IPA ($1:3$) for $45\,s$. To stop the development, the sample was rinsed with pure IPA. For the metallisation an e-beam evaporator was used to deposit $35\,nm$ of gold at a low rate of $1\,\angstrom /s$. The lift-off was performed in  warm acetone, leaving behind high-quality nanostructures with feature sizes on the order of $10\,nm$.
\hfill\break

\textbf{Simulations.} We simulated the scattering cross-section of the bowtie nanoantenna using a commercially available finite difference time domain solver (Lumerical Solutions, Inc., FDTD solutions, version: 8.11.387). We used a three-dimensional simulation cell that is terminated by perfectly matched layers. The bowtie was modelled using the extruded $N$-sided equilateral polygon with rounded corners that is provided on the Lumerical homepage (\url{https://kb.lumerical.com/en/ref_sim_obj_creating_rounded_corners.html}). At the centre of the simulation cell, i.e. around the bowtie feed-gap region, we used a mesh size of $2\,nm$, whereas in the outer regions the value was set to $4\,nm$. The computation of the scattering cross-section is based on the Mie scattering tutorial that can be found on the Lumerical hompage (\url{https://kb.lumerical.com/en/sp_fluorescence_enhancement.html}). Consequently, we excited the structures using a total field scattered field (TFSF) source and a finite-difference time-domain (FDTD) scattered field monitor to compute the scattering cross-section. The simulation principle of the hybrid system is fully based on the "Fluorescence Enhancement" tutoarial that is also published on the Lumerical homepage (\url{https://kb.lumerical.com/en/sp_fluorescence_enhancement.html}).
\hfill\break

\textbf{Optical measurements.} For differential reflection spectroscopy, a white light continuum laser (Fianium WhiteLaseTM micro) is focused onto individual bowtie nanoantennas using an $100\times$ microscope objective (Mitutoyo) with numerical aperture $NA=0.5$. Excitation power density and linear polarisation are adjusted with a continuous power attenuator wheel and a broadband linear polariser (Thorlabs, $LBVIS100-MP2$). The reflected signal from the bowtie nanoantennas is collected via the same objective, directed via a multimode optical fibre to a  $0.5\,m$ imaging spectrometer (Princeton Instruments Acton SP2500i, grating: $300\,l/mm$) and analysed with either a Si-CCD camera (Princeton Instruments Spec-10) or InGaAs diode array (Princeton Instruments, OMA V). The differential reflection spectrum is obtained from reflectivity spectra recorded at the position of the bowtie nanoantenna $R_{on}$ and at a position on the bare substrate as a reference $R_0$, by calculating $(R_{on}-R_0)/R_0$ \cite{Kaniber2016}.

Photoluminescence spectroscopy was carried out in a dip-stick cryostat at liquid Helium temperature $T=4.2\,K$. We used a continuous wave titanium-sapphire laser (Spectra-Physics Model 3900) with an excitation energy at $1.455\,eV$ to excite the quantum dots via the two-dimensional wetting layer. In addition, a defocused HeNe laser emitting above-bandgap at an emission energy of $1.962\,eV$ was used to stabilise the fluctuating charge environment around the quantum dot. Hereby, we used an excitation power density low enough to guarantee that no considerable photoluminescence signal from the quantum dots is generated, but high enough to reduced the quantum dot linewidth. The excitation power density was continuously monitored and stabilised to less than $0.1\,\%$ fluctuations and a uniform input polarisation was achieved by a fixed linear polariser (Thorlabs, $LBVIS100-MP2$). The excitation was focused onto the sample by a home-built low-temperature, achromatic microscope objective with $NA=0.7$. In the excitation path, an additional half wave plate (Thorlabs, AHWP10M-980) was introduced that allowed for tuning of the excitation polarisation. The emitted photoluminescence signal from individual quantum dots was collected via the same objective and analysed subsequently with a linear polariser (Thorlabs, $LBVIS100-MP2$), before guided via a polarisation-maintaining optical fibre to a spectrometer (Jobin Yvon Triax 550, grating $600\,l/mm$) equipped with a Si-CCD camera (Jobin Yvon Symphony CCD). For time-resolved measurements, the spectrometer was used as a monochromator to spectrally filter the emission of a single quantum dot before it is detected by a single photon avalanche detector (EG\&G Photon Counting Module SPCM-200) and analysed with time tagging electronics (Picoquant TimeHarp 220).

%
%
%
\hfill\break

\paragraph{Acknowledgements:} We gratefully acknowledge financial support from the DFG via SFB 631, the German Excellence Initiative via NIM, as well as support from the Technische Universit\"at M\"unchen (TUM) - Institute for Advanced Study, funded by the German Excellence Initiative and the TUM International Graduate School of Science and Engineering (IGSSE). A.A.L. acknowledges financial support of RFBR via the project no. 16-37-60075 mol\_a\_dk and the fellowship of Russian President SP-3014.2016.3.

%
%
%
\hfill\break

\paragraph{Author contributions} K.S. and A.A.L. designed and fabricated the sample. K.S. design and built the low-temperature dip-stick cryostat. K.S. and A.R. performed the structural measurements. A.A.L., K.S., A.R., and M. S. performed the optical spectroscopy. K.S. performed the numerical simulations. K.S., M.K., A.A.L., and S.P.M. analysed the data. A.K.B, A.I.T., and S.P.M were responsible for the design and growth of the semiconductor heterostructure. M.K. wrote the paper with contributions from all other authors. M.K proposed, initiated, coordinated and supervised the project. All authors discussed the results and reviewed the manuscript.

%
%
\FloatBarrier
\bibliographystyle{apsrev}
\bibliography{Bibliography_160215_QD-BT_v1}

\end{document}